\documentclass[epsfig,graphics,twocolumn,floatfix,mathbbm,pra]{revtex4}
\usepackage{graphicx}

\begin{document}

\title[a]{Multimode quantum memory based on atomic frequency combs}%

\author{Mikael Afzelius}
\email{mikael.afzelius@unige.ch}
\author{Christoph Simon}
\author{Hugues de Riedmatten}
\author{Nicolas Gisin}
\address{Group of Applied Physics, University of Geneva, CH-1211 Geneva 4, Switzerland}%

\date{\today}

\begin{abstract}
An efficient multi-mode quantum memory is a crucial resource for long-distance quantum communication based on quantum repeaters. We propose a
quantum memory based on spectral shaping of an inhomogeneously broadened optical transition into an atomic frequency comb (AFC). The spectral
width of the AFC allows efficient storage of multiple temporal modes, without the need to increase the absorption depth of the storage material,
in contrast to previously known quantum memories. Efficient readout is possible thanks to rephasing of the atomic dipoles due to the AFC
structure. Long-time storage and on-demand readout is achieved by use of spin-states in a lambda-type configuration. We show that an AFC quantum
memory realized in solids doped with rare-earth-metal ions could store hundreds of modes or more with close to unit efficiency, for material
parameters achievable today.
\end{abstract}

\maketitle

\section{INTRODUCTION}

The distribution of entanglement between remote locations is critical for future long-distance quantum networks and extended tests of quantum
non-locality. It is likely to rely on quantum repeaters \cite{Briegel1998,Duan2001}, which require quantum memories that can store entanglement
between distant network nodes \cite{Chou2007,Yuan2008}. Recent experimental achievements in quantum state storage
\cite{Julsgaard2004,Chaneliere2005,Eisaman2005,Choi2008} demonstrate that currently investigated QMs can store a single mode. Yet, long-distance
quantum repeaters having QMs only capable of storing one mode would only generate very limited entanglement generation rates
\cite{Sangouard2008a}. To achieve useful rates some way of multiplexing the QM will be required \cite{Collins2007,Simon2007}. By using time
\cite{Simon2007}, spatial \cite{Collins2007,surmacz-2007,Vasilyev2008}, or frequency multiplexing to store single photons in many modes $N$, the
entanglement generation rate can be increased by a corresponding factor $N$ \cite{Simon2007}. Here we consider multi-mode QMs capable of storing
$N$ temporally distinguishable modes, which is a natural form of sending information also used in today's telecommunications networks. Time
multiplexing is extremely challenging using current QM protocols such as stopped light based on Electromagnetically Induced Transparency (EIT)
\cite{Fleischhauer2000}, photon echoes based on Controlled Reversible Inhomogeneous Broadening (CRIB) \cite{Moiseev2001,Kraus2006} or
off-resonant Raman interactions \cite{Nunn2007}. For a single mode, $N = 1$, the efficiency as a function of optical absorption depth $d$ is
similar for EIT, CRIB, and Raman-type memories \cite{Gorshkov2007a,Gorshkov2008}. For multi-mode storage using EIT and Raman, however, $N$
scales as $\sqrt{d}$ \cite{Lukin2003,Nunn2008a}, which severely limits the number of potential modes in a practical experiment, while CRIB
offers a better scaling since $N \sim d$ \cite{Simon2007,Nunn2008a}. Still, even using CRIB, efficient multi-mode storage requires extreme
absorption depths that are currently out of reach.

In this article we present a QM, where the number of potential modes is essentially independent of the optical depth. As we will show, efficient
storage of 100s of modes or more is then feasible with realistic material parameters. The QM is based on absorption of light by an ensemble of
atoms on an optical transition which is inhomogeneously broadened. Ensembles are in general very attractive as QMs due to strong collective
enhancement of the light-matter coupling \cite{Duan2001}. Storage of single photons using EIT (stopped light) has been demonstrated with cold
alkali atoms \cite{Chaneliere2005,Choi2008} which can be treated as homogeneous ensembles of identical atoms. Here we will instead consider
ensembles of rare-earth-metal (RE) ions in solids, which are inhomogeneously broadened. Stopped light with storage times up to 1 s has been
demonstrated in RE-doped solids \cite{Turukhin2001,Longdell2005}, where an approximate homogeneous ensemble was created by spectrally isolating
a narrow absorption peak through optical pumping. In contrast to this approach, the quantum memory proposed here uses the inhomogeneous
broadening as a resource in order to achieve better multi-mode performance. For that one needs to coherently control the dephasing which is
caused by the inhomogeneous frequency distribution of the atoms. The key feature of our proposal is to achieve this control by a specific
shaping of this distribution into an atomic frequency comb.

The article is organized in the following way. In Section II we give an overview of the proposal using a simplified physical picture. In Section
III we show results from an analytical analysis of the physics of atomic frequency combs, which is further detailed in the Appendix. In Section
IV the multi-mode storage capacity is discussed, and in Sec. V we discuss implementation of the quantum memory in rare-earth-ion-doped solids.
Conclusions are given in Sec. VI.

\section{ATOMIC FREQUENCY COMBS}

\begin{figure}
    \centering
    \includegraphics[width=.45\textwidth]{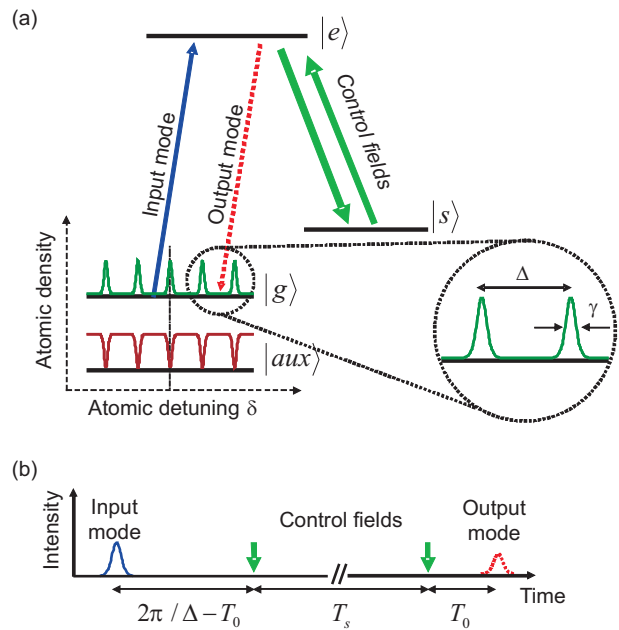}
    \caption{(Color online) The principles of the proposed atomic frequency comb (AFC) quantum memory.
    (a) An inhomogeneously broadened optical transition $|g\rangle -
|e\rangle$ is shaped into an AFC by frequency-selective optical pumping to the $|aux\rangle$ level. The peaks in the AFC have width (FWHM)
$\gamma$ and are separated by $\Delta$, where we define the comb finesse as $F = \Delta/\gamma$. (b) The input mode is completely absorbed and
coherently excites the AFC modes, which will dephase and then rephase after a time $2\pi/\Delta$, resulting in a photon-echo type coherent
emission. A pair of control fields on $|e\rangle - |s\rangle$ allow for long-time storage as a collective spin wave in $|s\rangle$, and
on-demand read-out after a storage time $T_s$.}
    \label{fig_prot}
\end{figure}

The storage material is an ensemble of atoms with an excited state $|e\rangle$ optically connected to two lower states $|g\rangle$ and
$|s\rangle$, see Fig. \ref{fig_prot}. The states $|g\rangle$ and $|s\rangle$ are typically ground-state hyperfine or Zeeman states. We assume
that the optical transition $|g\rangle - |e\rangle$ has a narrow homogeneous linewidth $\gamma_h$ and a large inhomogeneous broadening
$\Gamma_{in}$ ($\Gamma_{in}/\gamma_h \gg 1$). This is the case for RE-doped solids that we will consider for physical implementation in Sec. V.
The $|g\rangle - |e\rangle$ transition is spectrally shaped such that the atomic density function consists of a series of narrow peaks spanning
a large frequency range (atomic frequency comb, or AFC). This can be done by frequency-selectively transferring atoms from $|g\rangle$ to a
metastable state $|aux\rangle$, for instance a third hyperfine state, through optical pumping techniques (see Sec. V). Note that the maximum
spectral width of the AFC is limited by the level spacings between the ground state levels.

We then consider a single-photon input field, having a spectral distribution $\gamma_p$ larger than the AFC peak separation $\Delta$, but
narrower than the total width of the AFC $\Gamma$, which is in resonance with the $|g\rangle - |e\rangle$ transition. If the atomic density
integrated over the photon bandwidth is high enough the photon can be totally absorbed by the AFC, although the spectral density of atoms is
concentrated to narrow peaks. This can be understood in terms of the Heisenberg energy-time uncertainty relation. For the time scale of the
absorption, which is of the order of the input pulse duration $\tau=1/\gamma_p$, the optical transition will have an uncertainty of the order of
$\gamma_p \gg \Delta$. This causes a spectral averaging of the sharp AFC structure into a smooth distribution, allowing for uniform absorption
over the photons bandwidth (see also Sec. III).

After absorption (at $t$ = 0) the photon is stored as a single excitation de-localized over all the atoms in the ensemble that are in resonance
with the photon. The state is described by a collective Dicke state \cite{Dicke1954}

\begin{equation}|\psi\rangle=\sum_{j=1}^N c_j e^{i\delta_j
t}e^{-ikz_j} |g_1\cdot\cdot\cdot e_j\cdot\cdot\cdot g_N \rangle, \label{Dicke}
\end{equation}

\noindent where $z_j$ is the position of atom $j$, $k$ is the wave-number of the light field (for simplicity, we only consider a single spatial
mode defined by the direction of propagation of the input field), $\delta_j$ the detuning of the atom with respect to the laser frequency and
the amplitudes $c_j$ depend on the frequency and on the spatial position of the particular atom $j$.

The collective state can be understood as a coherent excitation of a large number of AFC modes by a single photon. These modes are initially (at
$t=0$) in phase with respect to the spatial mode $k$. But the collective state will rapidly dephase into a non-collective state that does not
lead to a strong collective emission, since each term acquires a phase $\exp(i\delta_jt)$ depending on the detuning $\delta_j$ of each excited
atom. If we consider an AFC having very sharp peaks, then the detunings $\delta_j$ are approximately a discrete set such that $\delta_j =
m_j\Delta$, where $m_j$ are integers. It follows that the collective state is re-established after a time $2\pi/\Delta$, which leads to a
coherent photon-echo \cite{Mossberg1979,Carlson1984,Mitsunaga1991} type re-emission in the forward direction. The efficiency of this process
(see Sec. III for details) can reach 54$\%$ in the forward direction (limited only by re-absorption). But if the re-emission is forced to
propagate in the backward direction, by a proper phase matching operation (see below), the process can reach 100$\%$ efficiency.

The process described so far only implements a QM with a fixed storage time. In order to allow for on-demand read-out of the stored field (which
is a necessary requirement for use in quantum repeaters) and long-term storage, the single collective excitation in $|e\rangle$ is transferred
to a ground state spin level $|s\rangle$. This can be done by applying an optical control field on $|e\rangle - |s\rangle$, for instance a short
$\pi$ pulse. The excitation is thereafter stored as a collective spin wave, which is also the basis of storage in EIT, CRIB, and off-resonant
Raman type memories based on ensembles of lambda atoms \cite{Fleischhauer2000,Moiseev2001,Nunn2007}. The spin wave allows for long-lived storage
since spin coherence lifetimes are generally longer than the optical coherence lifetimes. Moreover, the spin transition does not have a comb
structure \cite{comment_comb_spintrans}, which means that the evolution of the phases in Eq. (\ref{Dicke}) will be locked during the storage in
the spin wave. If we assume that the spin transition is completely homogeneous the storage time is only limited by the coherence lifetime of the
spin wave. If there is inhomogeneous broadening, spin echo techniques can be used to exploit the homogeneous coherence lifetime of the system
(see Sec. V). To retrieve the field a second counter-propagating control field is applied after some time $T_s$. These operations will result in
a backward propagating output mode after a total time $T_s+2\pi/\Delta$. As we will show below, the efficiency of the storage for backward
retrieval can theoretically approach unity for high optical depths.

\section{ANALYTICAL TREATMENT}

We will now discuss some important features of the AFC and give quantitative formulas relating the input field to the output field as a function
of the AFC parameters. This can be used to calculate, for instance, storage efficiency. To this end we have used an approximate, linearized
Maxwell-Bloch (MB) model. A linearized model can be used since the population in the excited state is negligible. For a single photon input
field at most one atom in the ensemble is excited. For multi-mode storage, the number of totally excited atoms should be much less than the
total number of atoms, which is the case for the number modes we consider and the RE-solids we discuss in Sec. V. The linearity of the
analytical model also implies the equivalence between classical and single-photon dynamics \cite{Sangouard2007,Nunn2007,Gorshkov2007a}. The
important features of the model is discussed below, while the details of the calculations are given in the appendix.

The matter part was assumed to be an ensemble of two-level atoms, including levels $|g\rangle$ and $|e\rangle$ in Fig. \ref{fig_prot}, and the
light field was treated as a one-dimensional spatial mode. This model can fully describe the mapping of light onto the atoms, the atomic
evolution including the inhomogeneous dephasing and rephasing of the AFC modes, and the re-emission of light. The model did not include
population relaxation ($T_1$) or decoherence ($T_2$). This is a valid approximation since the peaks in the AFC are assumed to be inhomogeneously
broadened. Then all relevant dephasing times are much faster than $T_1$ and $T_2$ (as in the example given in Sec. V). In practice, however, the width of the individual peaks will be limited by the homogeneous linewidth $\gamma_h=1/(\pi T_2)$. The two
counter-propagating control fields were modelled by transferring the forward atomic mode, associated with the forward propagating optical mode,
to a backward atomic mode \cite{Sangouard2007}. This achieves the phase matching operation that results in a backward propagating optical mode.
Clearly this approach assumes that the transfer efficiency of the control pulses is unity and that the spin transition $|g\rangle-|s\rangle$ is
decoherence free. We thus only analytically treat the physics of the atomic frequency comb, which is the novel feature of this protocol as
compared to previous quantum memory proposals \cite{Fleischhauer2000,Moiseev2001,Nunn2007}. The transfer efficiency and spin decoherence could,
however, be straightforwardly included with appropriate loss factors. The equations of motion describing the light-matter dynamics is given in the appendix (see Eqs (\ref{em_1})-(\ref{em_4})). In addition to the analytical calculations, we also show results from numerical solutions of
the MB model.

The atomic spectral distribution $n(\delta)$ is a frequency comb described by a series of Gaussian functions characterized by the peak
separation $\Delta$ and peak width $\tilde{\gamma}$, with an overall comb width of $\Gamma$

\begin{equation}
n(\delta) \propto e^{-\delta^2/(2 \Gamma^2)} \sum \limits_{j=-\infty}^{\infty} e^{-(\delta-j \Delta)^2/(2 \tilde{\gamma}^2)} \label{n}
\end{equation}

\noindent The full-width at half-maximum (FWHM) peak width is then $\gamma= \sqrt{8 \ln 2} \tilde{\gamma}$, and we define the finesse $F$ of the
comb as $F=\Delta/\gamma$. We are interested in the regime where we have a significant number of well-separated peaks within the overall
characteristic width of the AFC. Moreover, we assume that the spectrum of the photon is narrower than the width of the AFC, but wider than the
peak separation, i.e. $\Gamma \gg \gamma_p \gg \Delta \gg \tilde{\gamma}$.

In order to have an efficient memory, the probability of absorption of the input mode must be sufficiently high. But to create a high finesse
AFC atoms are removed, by optical pumping, from the natural inhomogeneous spectrum. This preparation step lowers the total density of atoms and
hence the probability of absorption. The analytical calculations show that the amplitude of the forward propagating mode $E_f(z,t)$ after the
sample of length $L$ is given by (see Eqs (\ref{ef})-(\ref{eff_d}))

\begin{equation} E_f(L,t)=E_f(0,t)
e^{-\frac{\tilde{\alpha L}}{2}}=E_f(0,t) e^{-\frac{\tilde{d}}{2}}, \label{eL}
\end{equation}

\noindent where the effective absorption depth $\tilde{d}=\tilde{\alpha}L$ can be expressed as

\begin{equation}
\tilde{d} = \frac{d}{F} \sqrt{\frac{\pi}{4\ln 2}} \approx \frac{d}{F}. \label{eq_d}
\end{equation}

The comb absorption is thus reduced by a factor of $1/F$, which is to be expected since the total density of atoms in the comb (integrated over
the frequency bandwidth of the photon) is reduced by a factor of $1/F$ in the preparation step. But if the peak absorption depth $d$ of the comb
is high enough, the total density of atoms in the comb can still be sufficiently high to absorb the input mode with close to 100$\%$
probability.

An interesting feature of the AFC is that it can absorb uniformly over the frequency spectrum of the input mode although the atomic frequency
comb has large "holes". This can be understood by considering that the absorption is an event well localized in time, of the order $\tau =
1/\gamma_p$. In that case the light-atom system reacts with a Fourier-limited resolution $\gamma_p$, which effectively smears out the structure
of the AFC to an essentially flat distribution (see discussion preceeding Eq. (\ref{ef_diff})). In other words the peak separation $\Delta$ must
be smaller than the photon bandwidth $\gamma_p$. This also implies that Eq. (\ref{eL}) is only valid for times $t \ll 2\pi/\Delta$, i.e. before
the appearance of the first emission due to rephasing of modes in the comb.

By applying the two counter-propagating control fields, the spatial phase pattern of the atomic polarization changes such that it couples with
the backward propagating field mode \cite{Moiseev2001,Kraus2006,Sangouard2007}. Under the assumption of unit transfer efficiency and no phase
decoherence during the transfer to or storage in $|s\rangle$, the atomic phases will evolve to a collective state coupled to the backward
propagating mode. This leads to emission in the backward mode. In the appendix (see Eq. (\ref{Eout})) we show that the output field amplitude is
a coherent sum over the amplitudes corresponding to all possible paths of the photon in the medium, where each path is characterized by a
particular turning point in the medium.  Even though there is absorption also along the path of the backward mode, a constructive interference
effect between all these possible paths leads to the output amplitude potentially being arbitrarily close in magnitude to the input amplitude
for large enough $d$. This interference effect is also present in the CRIB quantum memory scheme for backward retrieval \cite{Sangouard2007}.
The analytical calculation leads to the following relation between the forward input field $E_{in}=E_f(z=0,t=0)$ and output backward field
$E_{out}=E_b(z=0,t=2\pi/\Delta)$ (see Eq. (\ref{Eout})),

\begin{equation}
E_{out} = -E_{in} e^{-i2\pi\frac{\Delta_0}{\Delta}} (1-e^{-\tilde{d}}) e^{-\frac{1}{F^2} \frac{\pi^2}{4 \ln 2}}. \label{eq_eff}
\end{equation}

\noindent The first factor represents a global phase shift that depends on the relative detuning $\Delta_0$ between the carrier frequency of the
light and the centre of the AFC. The second factor describes the coupling between the input and output fields, which increases with increasing
effective absorption depth, Eq. (\ref{eq_d}), whereas the third factor takes into account dephasing due to the finite width of the peaks.

The dephasing factor can be understood by considering the dephasing during the storage in the optically excited state $|e\rangle$. In the
qualitative description in Sec. II we assumed that each peak was a delta function, which leads to complete rephasing of the collective state Eq.
(\ref{Dicke}). Clearly finite Gaussian peaks will not lead to complete rephasing, which lowers the probability of re-emission. The decay of the
re-emitted field amplitude is given by the Fourier-transform of a peak in the comb (see Eq. (\ref{ntilde}) and discussion preceding Eq.
(\ref{eb-simple})), in analogy with the role of the initial peak shape in the CRIB quantum memory protocol \cite{Alexander2006,Sangouard2007}.
In the case of Gaussian peaks, the decay is given by $e^{-t^2\tilde{\gamma}^2/2}$, where $t=2\pi/\Delta$ is the storage time in the excited
state. This factor can be expressed in terms of the finesse $F$, which leads to the last term in Eq. (\ref{eq_eff}). To reduce this dephasing
one should thus maximize the comb finesse $F$.

\begin{figure}
    \centering
    \includegraphics[width=.50\textwidth]{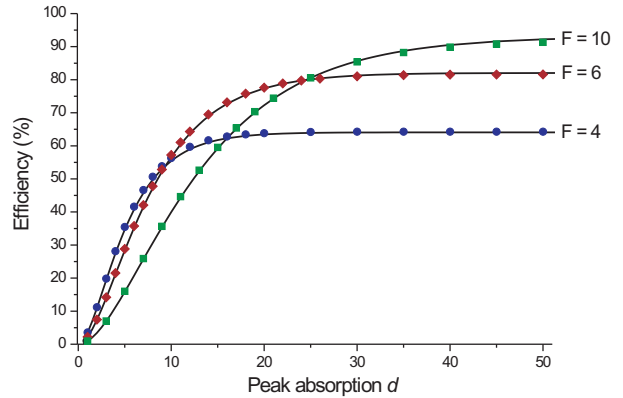}
    \caption{(Color online) The storage efficiency is shown as a function of peak absorption depth $d$ for AFC finesse
    values of $F$ = 4, 6 and 10. The solid lines represent the efficiency calculated using the analytical
    result, Eq. (\ref{eq_eff}), whereas the symbols represent the numerical calculation. As an example,
    for $d $ = 40 and $F$ = 10, an efficiency of 90$\%$ can be achieved. Even for an absorption depth of 25
    and a modest finesse of 6, the efficiency reaches 80$\%$.}
    \label{fig_eff}
\end{figure}

\begin{figure}
    \centering
    \includegraphics[width=.50\textwidth]{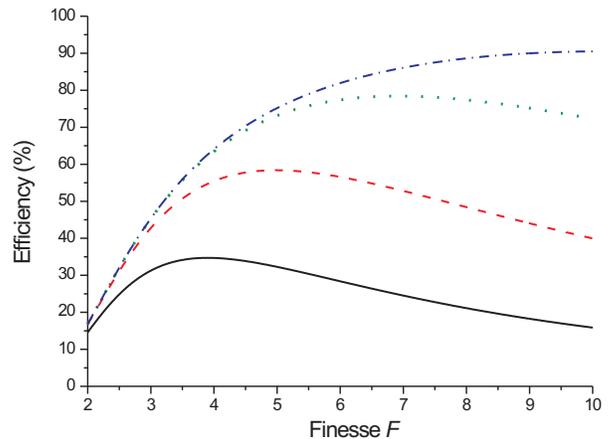}
    \caption{(Color online) Efficiency as a function of comb finesse $F$ for
    peak absorption depths $d=5$ (black solid line), $d=10$ (red dashed line), $d=20$ (green dotted line) and $d=40$ (blue dashed-dotted line).
    The curves were calculated using Eq. (\ref{eq_eff}).}
    \label{fig_eff_F}
\end{figure}

The efficiency, defined as $\eta=|E_{out}|^2/|E_{in}|^2$, is plotted as a function of optical depth for different comb finesses in Fig.
\ref{fig_eff}. For a moderate finesse of $F=4$, an efficiency greater than 50$\%$ can be achieved for $d=10$. For a high finesse, $F > 10$, the
efficiency can be close to 100$\%$ for large enough $d$. From an experimental point of view $d$ is often a fixed parameter. The storage
efficiency can then be maximized for the given material by optimizing $F$ using Eq. (\ref{eq_eff}) (see Fig. \ref{fig_eff_F}). This optimum in
the choice of $F$ is due to a trade-off between absorption probability (favoring small $F$) and low dephasing during storage (favoring large
$F$).

We will also briefly discuss the possibility of re-emission in the forward direction. This will occur if no control fields are applied or if the
control fields are applied co-propagating. As is shown in the Appendix (see Eq. (\ref{E_out_fw})), this re-emission is limited in efficiency to
54$\%$ for an optimal optical depth of $\tilde{d}=2$. The optimum represents a trade-off between strong light-matter coupling (favoring high
optical depth) and re-absorption of the re-emitted light as it propagates through the sample (favoring low optical depth).

\section{MULTI-MODE STORAGE CAPACITY}

The excellent multi-mode properties of the AFC based QM can be understood qualitatively as follows. We imagine that the input is a series of
temporally distinguishable modes containing, with a certain probability, a single photon in each mode. The storage efficiency is uniform over all the modes, since each mode spends the same time in the memory. Note that the effect of spin decoherence is also the same on each mode. If the control fields are applied as late
as possible, the total duration of the input can be close to $T=2\pi/\Delta$, the time it takes to rephase the first mode that was absorbed. The
shortest duration $\tau$ of one mode, however, is only limited by the total frequency bandwidth $\tau\sim 1/(N_p\Delta)$, where $N_p$ is the
number of peaks in the comb. The number of modes $N=T/\tau\sim N_p$ is then proportional to $N_p$, such that adding more peaks in the AFC allows
one to store more modes. Thus, in materials with inherently large inhomogeneous broadenings (see Sec. V) one can increase the multi-mode
capacity by making use of a wider range of the inhomogeneous broadening, without having to increase the peak absorption depth by, for instance,
increasing the atomic density in the sample.

We can make a more detailed calculation of the multi-mode capacity by assuming that the input modes have Gaussian temporal shapes, where each
mode occupies a temporal slot of duration $\tau$. It is important that the modes do not overlap too much, which in the end would produce errors.
The amount of overlap that can be accepted depends on the particular application. In Ref. \cite{Simon2007} an analysis of the error due to
overlap was carried out in the context of a quantum repeater protocol in which the AFC QM could be used. There it was shown that $\tau$ should
be chosen as $\tau=12\pi/\Gamma$, where $\Gamma$ is the total memory bandwidth (note that here $\Gamma$ is expressed in radians), where the
width $\Gamma$ is taken to be that of a square distribution, as in Ref. \cite{Simon2007}. This also guarantees that the spectral content of the input modes can be fitted inside the AFC, which in turn results in negligible temporal distortion of the output modes. The width $\Gamma$ can be related to the number of AFC
peaks $N_p = \Gamma/\Delta$. The total duration $T$ of the input is $T=2\pi/\Delta-T_0$, where $T_0$ is the time necessary for applying the two
control fields. In the limit where $T_0\ll2\pi/\Delta$, we have $T\approx2\pi/\Delta$. The number of total temporal modes then amounts to $N =
T/\tau\approx N_p/6$.

The number of peaks that can be created will depend strongly on the inhomogeneous $\Gamma_{in}$ and homogeneous $\gamma_h$ linewidths of the
material. The width of each peak cannot be narrower than the homogeneous linewidth, i.e. $\gamma > \gamma_h$. Since an efficient memory
($>90\%$) requires a finesse $F>10$, it follows that $\Delta > 10\gamma_h$. The number of peaks will depend on the ratio of the maximum AFC
bandwidth $\Gamma$ to the peak separation $\Delta$. The bandwidth will ultimately be limited by the inhomogeneous broadening of the material. In
most materials, however, it will instead be limited by hyperfine or Zeeman transition spacings in order to avoid exciting multiple transitions
that would induce quantum beat effects. In Section V we will give a detailed example where the number of peaks is of the  order of hundreds with
known material parameters.

The AFC QM dramatically reduces the necessary absorption depth for multi-mode storage as compared to CRIB or EIT QMs, since the multi-mode
capacity is independent of the peak absorption depth of the material. For CRIB, $d = 30N$ is required for storing $N$ modes with an average
efficiency of 90$\%$ \cite{Simon2007}, so for instance $N$ = 100 requires $d$ = 3000. In the case of EIT even larger $d$ would be necessary due
to the worse scaling as mentioned in the introduction. The AFC only requires $d = 40$ to reach the same efficiency, independently of $N$. This
makes efficient multi-mode QMs more realistic with current material properties, as we will discuss in the next Section.

\section{IMPLEMENTATION USING RARE-EARTH-ION-DOPED SOLIDS}

As storage material we here consider RE-doped crystals, which have inherently large inhomogeneous broadenings (0.1 - 10 GHz) and very narrow
homogenenous linewidths (0.1-100 kHz) at low temperatures (typically less than 4 K) \cite{Macfarlane2002}. Note that the typical homogeneous
linewidths correspond to optical coherence times ranging from $\mu$s to ms. Still RE ensembles have relatively high absorption coefficients at
normal doping concentrations (10-100 ppm), typically in the range 1-40 cm$^{-1}$. RE crystals are therefore good candidates for the high
resolution spectral shaping we are discussing here. Ref. \cite{Macfarlane2002} contains an extensive review of optical properties of different
RE-doped crystals and glasses.

The creation of the periodic narrow structure is the crucial enabling step for the AFC QM. Narrow, absorbing peaks ($<$ 50 kHz) have been
isolated within the inhomogeneous broadening by optical pumping (or spectral hole burning) \cite{Seze2003,Nilsson2004,Hetet2008,Lauritzen2008}.
In all stopped-light \cite{Turukhin2001,Longdell2005} and CRIB experiments \cite{Alexander2006,Hetet2008} in RE solids to this date, a narrow
absorption peak within a larger transmission hole has been created in order to have an approximate homogeneous distribution of atoms. Briefly, a
large transmission hole is first created by spectral hole burning on the optical transition while sweeping the laser frequency. The pumped atoms
are stored in another hyperfine level with long population lifetime. A narrow ensemble of atoms is then created in the hole by repumping atoms
from the storage state to the initial state using a highly coherent laser source \cite{Rippe2005}. By repumping at different frequencies this
method can be extended in order to create a series of peaks. One can also use other techniques in order to achieve frequency-selective optical
pumping. For instance, one can use pairs of coherent pulses where each pair coherently transfers atoms to the excited state via a Ramsey
interference \cite{Hesselink1979}. Numerical simulations show that by accumulating many pairs, narrow, periodic structures can be created. This
time-domain approach is interesting because in principle a time sequence can be created which has the appropriate frequency spectrum for
creating an AFC with a certain finesse. Moreover a time-domain approach can be used for creating more complicated structures in frequency space,
for instance multiple AFCs having different periodicity.

The spin coherence times have the potential of being significantly longer than the optical ones since it is free of decoherence due to
spontaneous emission. In Pr$^{3+}$-doped Y$_2$SiO$_5$ crystals, one of the most studied RE-doped material, the spin coherence time is 500 $\mu$s
at zero magnetic field \cite{Ham1998}, which can be extended to 82 ms by applying an orientation-specific magnetic field \cite{Fraval2004}.
Fraval et al. \cite{Fraval2005} has also shown that it is possible to use dynamic decoupling methods developed in NMR spectroscopy to further
extend the coherence time to 30 s. Yet another example is Eu$^{3+}$:Y$_2$SiO$_5$ where 15 ms spin coherence time was measured at zero field
\cite{Alexander2007}, and Tm$^{3+}$:YAG where 300 $\mu$s was obtained in a magnetic field \cite{Louchet2008}. Efficient transfer of population
between the optically excited state and ground hyperfine states has been demonstrated in Pr$^{3+}$:Y$_2$SiO$_5$, and manipulation of coherence
between the optically excited state and the two spin ground states has recently been demonstrated in a state tomography experiment of an
ensemble-based spin qubit in Pr$^{3+}$:Y$_2$SiO$_5$ \cite{Rippe2008}. This shows the potential of spin states for long-term storage states for
quantum information.

We finally give a more detailed example based on Europium-doped Y$_2$SiO$_5$ crystals, to show the potential for efficient multi-mode storage
using an AFC based memory. Europium ions absorb at 580 nm and have an appropriate energy structure with three ground state levels and an
optically excited state \cite{Nilsson2002}, as well as excellent optical \cite{Koenz2003} and hyperfine \cite{Alexander2007} coherence times.
Given the very narrow homogeneous linewidth in this material (122 Hz \cite{Koenz2003}), one could realistically create an AFC with $\gamma$ = 2
kHz and $\Delta$ = 20 kHz (i.e. $F = 10$). The total AFC bandwidth we take to be 12 MHz, limited by hyperfine transition spacings, which results
in $N_p$ = 600 peaks. The maximum number of modes is then roughly $N = N_p/6 \approx 100$. The absorption coefficient is about 3-4 cm$^{-1}$
\cite{Koenz2003}, so that $d = 40$ is within technical reach today using a multi-pass arrangement. Hence, one could then store 100 temporal
modes in one single QM with an efficiency of 90$\%$, which we verified by numerically simulating storage of 100 modes using the parameters
above. We note that the efficiency stated above relates to the physics of the atomic frequency comb (see the model assumptions in sec. III). A
complete storage experiment including transfer to the spin state will be affected by the limitation in spin coherence time, for
Eu$^{3+}$:Y$_2$SiO$_5$ $T_2^{spin}$=36 ms \cite{Alexander2007} has been observed so far, and the transfer efficiency of the control fields (cf.
discussion above).

We conclude this section by observing that less studied RE-doped crystals could potentially store more modes due to larger hyperfine state
separations, for instance Erbium-doped crystals absorbing at the telecommunication wavelength 1.5 $\mu$m or Neodymium-doped crystals absorbing
at 880 nm.

\section{CONCLUSIONS}

We have presented a protocol for a quantum memory that can store efficiently a large number of temporal modes in a single atomic ensemble by
creating an atomic frequency comb. In materials with large optical inhomogeneous broadenings and narrow homogeneous broadenings this allows for
high multi-mode capacity at moderate optical absorption depths. We argue that RE-doped materials have the necessary properties for implementing
the proposed protocol. The inherent multi-mode property of the atomic frequency comb quantum memory promises to speed up the entanglement
generation rates in long-distance quantum networks by several orders of magnitude.\\
\indent The authors acknowledge useful discussions with Imam Usmani, Nicolas Sangouard, Thierry Chaneli\`{e}re, Jean-Louis Le Gou\"{e}t and
Stefan Kr\"{o}ll. This work was supported by the Swiss NCCR Quantum Photonics and by the European Commission under the Integrated Project Qubit
Applications (QAP).

\textit{Note added}. Since this paper was submitted for review, we have experimentally demonstrated a light-matter interface at the single
photon level \cite{deRiedmatten2008} using an atomic frequency comb.

\appendix
\section{ANALYTICAL CALCULATIONS}
\subsection{Equations of motion}

In this appendix we present the analytical calculations leading to the quantitative formulas in the Article. The basic equations describing the
dynamics of the light-atoms system are the same as in section II of Ref. \cite{Sangouard2007}. Both the field and the atomic polarization are
described by (slowly varying) forward and backward modes, denoted $E_f(z,t)$ and $E_b(z,t)$ for the field and $\sigma_f(z,t;\delta)$ and
$\sigma_b(z,t;\delta)$ for the atomic polarization corresponding to atoms with resonance frequency $\delta$. They satisfy the following
equations, cf. Eqs. (5a-b) and (6a-b) of Ref. \cite{Sangouard2007}.
\begin{eqnarray}
\frac{\partial}{\partial t} \sigma_f(z,t;\delta)=-i\delta
\sigma_f(z,t;\delta)+i\wp E_f(z,t) \label{em_1} \\
(\frac{\partial}{\partial t}+c \frac{\partial}{\partial z}) E_f(z,t)=i \tilde{\wp} \int \limits_{-\infty}^{\infty} d\delta
n(\delta) \sigma_f(z,t;\delta) \label{em_2}\\
\frac{\partial}{\partial t} \sigma_b(z,t;\delta)=-i\delta
\sigma_b(z,t;\delta)+i\wp E_b(z,t) \label{em_3} \\
(\frac{\partial}{\partial t}-c \frac{\partial}{\partial z}) E_b(z,t)=i \tilde{\wp} \int \limits_{-\infty}^{\infty} d\delta n(\delta)
\sigma_b(z,t;\delta). \label{em_4}
\end{eqnarray}
Here $\wp$ is the dipole moment of the atomic transition, and $\tilde{\wp}=g_0^2  \wp$, with $g_0=\sqrt{\omega_0/(2\epsilon_0 V)}$, $\omega_0$
is the optical frequency, and $V$ is the quantization volume; $n(\delta)$ is the atomic spectral distribution, with $\int
\limits_{-\infty}^{\infty} d\delta n(\delta)=N_a$, where $N_a$ is the total number of atoms. See Ref. \cite{Sangouard2007} for more details.
Note that $n(\delta)=N_a G(\delta)$ and $\tilde{\wp}=\beta/N_a$ in the notation of Ref. \cite{Sangouard2007}.

\subsection{Atomic spectral distribution and its Fourier transform}
Let us suppose that
\begin{equation}
n(\delta) \propto e^{-\delta^2/(2 \Gamma^2)} \sum \limits_{n=-\infty}^{\infty} e^{-(\delta-n \Delta)^2/(2 \tilde{\gamma}^2)},
\end{equation}
i.e. the atomic spectral distribution is a frequency comb with overall width $\Gamma$, peak separation $\Delta$ and individual peak width
$\tilde{\gamma}$. Note that in the Article we talk about the Full-Width at Half-Maximum (FWHM) peak width $\gamma= \sqrt{8 \ln 2}
\tilde{\gamma}$.

We are interested in the regime where $\Gamma \gg \Delta \gg \tilde{\gamma}$, i.e. we have a significant number of well-separated peaks within
the overall characteristic width. We define the finesse $F$ of the atomic distribution as the ratio of the peak separation over the FWHM of the
individual peak, which gives $F=\Delta/\gamma=\Delta/(\sqrt{8 \ln 2}\tilde{\gamma})$.

In the following the Fourier transform $\tilde{n}(t)$ of $n(\delta)$ will play an important role. One finds
\begin{equation}
\tilde{n}(t) \propto e^{-t^2 \tilde{\gamma}^2/2} \sum \limits_{n=-\infty}^{\infty} e^{-(t-n 2 \pi/\Delta)^2 \Gamma^2/2}, \label{ntilde}
\end{equation}
which is also a series of peaks, with overall temporal width $1/\tilde{\gamma}$, peak separation $2 \pi/\Delta$ and individual peak width
$1/\Gamma$. Under the above assumptions, the peaks are again numerous (within the characteristic overall width) and well-separated. To
summarize, both $n(\delta)$ and $\tilde{n}(t)$ are pulse trains, with inversely proportional separations between the individual pulses, and the
{\it global} shape of $n(\delta)$ determines the {\it local} shape of $\tilde{n}(t)$ and vice versa.

\subsection{Absorption}

The absorption process is described by the forward equations (\ref{em_1}-\ref{em_2}) coupling $E_f(z,t)$ and $\sigma_f(z,t;\delta)$. Plugging the solution of Eq.
(\ref{em_1}) into Eq. (\ref{em_2}) one has the following equation for the field
\begin{eqnarray}
(\frac{\partial}{\partial t}+c \frac{\partial}{\partial z}) E_f(z,t)=-\wp \tilde{\wp} \int \limits_{-\infty}^{\infty} d\delta n(\delta) e^{-i
\delta t}\times\nonumber\\ \int \limits_{-\infty}^t dt' e^{i \delta t'} E_f(z,t') \label{abs}.
\end{eqnarray}
This can be rewritten as
\begin{equation}
(\frac{\partial}{\partial t}+c \frac{\partial}{\partial z}) E_f(z,t)=-\wp \tilde{\wp} \int \limits_{-\infty}^t dt' \tilde{n}(t-t') E_f(z,t')
\label{abs-ft}
\end{equation}
with
\begin{equation}
\tilde{n}(t)=\int \limits_{-\infty}^{\infty} d\delta n(\delta) e^{-i \delta t}.
\end{equation}
Let us consider an incoming light pulse of temporal length $\tau$. Let us note right away that we are here considering the regime $\tau \gg
L/c$, where $L$ is the length of the medium. This allows us to neglect temporal retardation effects, i.e. we can neglect the temporal derivative
in Eqs. (\ref{abs}) and (\ref{abs-ft}). The absorption process then has a duration that is essentially given by $\tau$. After this time the
field is very close to zero, until the coherent photon-echo type re-emission at $t=2 \pi/\Delta$. If we are interested only in the absorption
process, we only need to consider values of $t-t'$ of order $\tau$ in Eq. (\ref{abs-ft}). Assuming $\tau \ll 2 \pi/\Delta$, i.e. a pulse
duration much shorter than the waiting time for the re-emission, only the central peak of $\tilde{n}$ from Eq. (\ref{ntilde}) will contribute,
i.e.
\begin{equation}
\tilde{n}(t-t') \propto e^{-(t-t')^2 \Gamma^2/2}
\end{equation}
for $t-t'$ of order $\tau$. Assuming that the overall width of the spectral distribution $\Gamma$ is much larger than the spectral width of the
incoming pulse, i.e. $\Gamma \tau \gg 1$, $\tilde{n}(t-t')$ will act essentially like a delta function in the integral in Eq. (\ref{abs-ft}),
giving an approximate equation of the form
\begin{equation}
\frac{\partial}{\partial z} E_f(z,t)=-\frac{\tilde{\alpha}}{2} E_f(z,t),\label{ef_diff}
\end{equation}
with a constant absorption coefficient $\tilde{\alpha}$. This is easily solved, giving \begin{equation} E_f(z,t)=E_f(0,t)
e^{-\frac{\tilde{\alpha}}{2} z}. \label{ef}
\end{equation} Concerning the absolute strength of the
absorption, one can show that under the above assumptions $\tilde{\alpha}=\alpha \frac{\sqrt{2 \pi} \tilde{\gamma}}{\Delta}$, where $\alpha$ is
the absorption coefficient corresponding to the (central) peaks of $n(\delta)$. This can be rewritten
\begin{equation}
\tilde{\alpha}=\frac{\alpha}{F} \sqrt{\frac{\pi}{4 \ln 2}}=1.064\frac{\alpha}{F}. \label{eff_d}
\end{equation}
Hence, for the absorption process, the AFC is indistinguishable from a smooth distribution that would be obtained by a spectral averaging that
does not resolve the individual peaks. Note that in the Article we frequently discuss the absorption depth $d$, here defined as $d = \alpha L$.
The effective absorption depth of the AFC is then $\tilde{d}=\tilde{\alpha}L$, such that equation (\ref{eff_d}) above corresponds to equation
(4) in the Article.

\subsection{Re-emission}

The pair of $\pi$ pulses, which are used for storing the excitations in a long-lived ground state and then bringing them back to the excited
state, also have the effect of changing the phase pattern of the atomic polarization from forward ($\sigma_f$) to backward ($\sigma_b$), such
that it now couples to the backward field $E_b$, cf. Ref. \cite{Sangouard2007}. We will see that this is essential for achieving potentially
perfect readout efficiency, in analogy with the case of CRIB memories \cite{Sangouard2007}. We assume that the time spent in the ground state
does not lead to any relative phases between the different frequency components (no inhomogeneous broadening in the ground state). From the
point of view of the equations of motion, one can thus treat the two $\pi$ pulses as being applied essentially simultaneously. We will refer to
the corresponding time as $t=0$. Absorption thus occurred at negative times, and re-emission will occur at positive times. The backward
polarization created at $t=0$ is
\begin{equation}
\sigma_b(z,0;\delta)=\sigma_f(z,0;\delta)=\int \limits_{-\infty}^{0} dt' e^{i \delta t'} E_f(z,t'), \label{sigmab-ini}
\end{equation}
where $\sigma_b$ ($\sigma_f$) is to be understood after (before) the two $\pi$ pulses. Eq. (\ref{sigmab-ini}) gives the initial condition for
Eqs. (3-4). We assume that the field $E_b$ is zero at $t=0$ (the absorption process is completed, if there was a transmitted field it has left
the medium). This gives the following equation for the backward field
\begin{eqnarray}
(\frac{\partial}{\partial t}-c\frac{\partial}{\partial z})E_b(z,t)=-\wp \tilde{\wp} \int \limits_{-\infty}^{0} dt' \tilde{n}(t-t')
E_f(z,t')\nonumber\\-\wp \tilde{\wp} \int
\limits_0^{t} dt' \tilde{n}(t-t') E_b(z,t'). \nonumber\\
\label{backward}
\end{eqnarray}
The first term on the right hand side is the source term which gives rise to the coherent re-emission, the second term describes the absorption
of the backward field in analogy with the case treated in the previous section, see Eq. (\ref{abs-ft}). The re-emission occurs because
$\tilde{n}(t-t')$ has a peak not only for $t-t'$ around zero, but also for $t-t'$ around $2 \pi/\Delta$. Note that the height of this peak is
reduced by a factor $e^{-(\frac{2 \pi}{\Delta})^2 \tilde{\gamma}^2/2}$ with respect to the central peak, cf. Eq. (\ref{ntilde}). Around the
corresponding point in time, one can apply arguments fully analogous to the ones in the previous subsection. Eq. (\ref{backward}) then
simplifies to
\begin{equation}
\frac{\partial}{\partial z} E_b(z,t)=\tilde{\alpha} e^{-\tilde{\gamma}^2 (\frac{2\pi}{\Delta})^2/2}
E_f(z,t-\frac{2\pi}{\Delta})+\frac{\tilde{\alpha}}{2} E_b(z,t). \label{eb-simple}
\end{equation}
The $E_f$ term has a factor $\tilde{\alpha}$ instead of $\frac{\tilde{\alpha}}{2}$ because the $t'$ integral in Eq. (\ref{backward}) is over all
of the effective delta function (given by the corresponding peak of $\tilde{n}$) for the $E_f$ (source) term, whereas it is only over one half
of the delta function for the $E_b$ (absorption) term, as it was in the previous subsection for the absorption of $E_f$.

The solution of Eq. (\ref{eb-simple}) is
\begin{equation}
E_b(z,t)= e^{-\tilde{\gamma}^2 (\frac{2\pi}{\Delta})^2/2}\int \limits_L^z dz'\tilde{\alpha} e^{\frac{\tilde{\alpha}}{2}(z-z')}
E_f(z',t-\frac{2\pi}{\Delta}).
\end{equation}
Note that $E_b$ propagates backwards, so the output field is at $z=0$. Using Eq. (\ref{ef}) one has
\begin{eqnarray}
E_b(0,t)&=&-E_f(0,t-\frac{2\pi}{\Delta}) e^{-\tilde{\gamma}^2 (\frac{2\pi}{\Delta})^2/2} \times \nonumber \\
&& \int \limits_0^L dz'\tilde{\alpha} e^{-\tilde{\alpha} z'/2} e^{-\tilde{\alpha} z'/2} \nonumber \\
&=&-E_f(0,t-\frac{2\pi}{\Delta}) e^{-\tilde{\gamma}^2 (\frac{2\pi}{\Delta})^2/2} (1-e^{-\tilde{\alpha} L}). \nonumber \\
\label{Eout}
\end{eqnarray}
It is interesting to note that the integral over $z'$ has the same form as in the case of the (backward) CRIB memory \cite{Sangouard2007}. It
corresponds to the coherent sum over all possible paths of the photon in the medium, where each path is characterized by its turning point $z'$.
Even though there is absorption along the path, there is a constructive interference effect which leads to the output amplitude potentially
being arbitrarily close in magnitude to the input amplitude (for large enough $\tilde{\alpha} L$).

This interference is not present if the output field is emitted in the forward direction, which would be the case if the two control fields were
not applied. Then one would have an integral
\begin{equation}
\int \limits_0^L dz' \tilde{\alpha} e^{-\tilde{\alpha} z'/2} e^{-\tilde{\alpha} (L-z')/2}=\tilde{\alpha}L e^{-\tilde{\alpha}L/2},
\label{E_out_fw}
\end{equation}
replacing the last factor in Eq. (\ref{Eout}). If one then calculates the efficiency, one finds that it can at most attain 54$\%$ for
$\tilde{\alpha}L = 2$, cf. \cite{Sangouard2007}.

For the backward propagating mode the achievable memory efficiency is given by, from Eq. (\ref{Eout}),
\begin{equation}
\eta=(1-e^{-\tilde{\alpha} L})^2 e^{-\tilde{\gamma}^2 (\frac{2\pi}{\Delta})^2},
\end{equation}
which can be rewritten in terms of the finesse $F$ and peak absorption $\alpha$ of the atomic distribution $n(\delta)$ as (cf. Eq. (5) in
Article)
\begin{eqnarray}
\eta &=& (1-e^{-\frac{\alpha L}{F} \sqrt{\frac{\pi}{4 \ln 2}}})^2 e^{-\frac{1}{F^2} \frac{\pi^2}{2 \ln 2}} \nonumber \\
&\approx& (1-e^{-\frac{\alpha L}{F}})^2 e^{-\frac{7}{F^2}}.
\end{eqnarray}
There is a trade-off between absorption (favoring small $F$) and dephasing (favoring large $F$), however, for large enough $\alpha L$ and
optimized $F$ the efficiency can be made arbitrarily close to 100$\%$ (cf. Figure 2 and 3 in Article).

\subsection{Phase due to frequency offset} Here we will comment on the phase of the output
field as a function of the relative frequency detuning between the carrier frequency and the frequency of the peak in the AFC (any of them), cf.
Eq. (5) in the Article. If the atomic distribution $n(\delta)$ is replaced by $n(\delta-\Delta_0)$, corresponding to a small frequency offset,
then the Fourier transform $\tilde{n}(t)$ acquires a phase factor $e^{i \Delta_0 t}$, which leads to a supplementary phase factor $e^{i
2\pi\frac{\Delta_0}{\Delta}}$ for the re-emitted field, cf. Eq. (\ref{Eout}). Note that this phase is modulus the peak separation $\Delta$.

\bibliographystyle{unsrt}

\begin{thebibliography}{10}
\bibitem{comment_comb_spintrans}
For the rare-earth-doped materials considered here the optical and spin inhomogeneous broadenings are generally uncorrelated. If the physical
system would contain correlations between these broadenings, however, there could be a comb structure also on the spin transition.
\end{thebibliography}

\end{document}